\documentclass[twocolumn,twocolappendix,tighten,times,astrosymb]{aastex7}

\usepackage{mathtools}
\usepackage[T1]{fontenc}
\usepackage{microtype}

\newcommand{\dep}{d_{e0}}              
\newcommand{\sigo}{\sigma_0}             
\newcommand{\sigin}{\sigma_{\text{in}}}  
\newcommand{\vp}{v_{\text{push}}}         
\newcommand{\bup}{B_{x\text{up}}}         
\newcommand{\td}{t_{\rm drive}}        
\newcommand{\tonset}{t_{\rm onset}}

\newcommand\bb[1]{\mbox{\boldmath{$#1$}}}

\newcommand\rmd{\mathrm{d}}
\newcommand{\vecE}{\bb{E}}
\newcommand{\vecB}{\bb{B}}

\shorttitle{\small Granier et al.}
\shortauthors{\small Granier et al.}

\begin{document}

\title{\large Driven Collisionless Reconnection of Force‐free Flux Tubes: From Onset to Coalescence}

\author[0000-0003-2841-8153]{Camille Granier}
\affiliation{Centre for mathematical Plasma Astrophysics, Department of Mathematics, KU Leuven,\\ Celestijnenlaan 200B, B-3001 Leuven, Belgium}
\email{camille.granier@kuleuven.be}

\author[0000-0002-5408-3046]{Daniel Gro\v{s}elj}
\affiliation{Centre for mathematical Plasma Astrophysics, Department of Mathematics, KU Leuven,\\ Celestijnenlaan 200B, B-3001 Leuven, Belgium}
\email{daniel.groselj@kuleuven.be}

\author[0000-0001-8822-8031]{Luca Comisso}
\affiliation{Department of Physics, Columbia University, New York, NY 10027, USA}
\affiliation{Department of Astronomy, Columbia University, New York, NY 10027, USA}
\email{luca.comisso@columbia.edu}

\author[0000-0002-7526-8154]{Fabio Bacchini}
\affiliation{Centre for mathematical Plasma Astrophysics, Department of Mathematics, KU Leuven,\\ Celestijnenlaan 200B, B-3001 Leuven, Belgium}
\affiliation{Royal Belgian Institute for Space Aeronomy, Solar-Terrestrial Centre of Excellence,\\ Ringlaan 3, 1180 Uccle, Belgium}
\email{fabio.bacchini@kuleuven.be}


\begin{abstract}
We investigate the onset of driven collisionless reconnection and plasmoid formation in a magnetically dominated pair plasma, using 2D Particle-in-Cell simulations.
Two force-free flux tubes of radius $R$ are initially pushed together with a prescribed velocity, forming a current sheet whose width shrinks until reconnection sets in. %
Even in our largest simulation with $R\approx 1600$ plasma skin depths, the sheet thickness at reconnection onset is comparable to the skin depth.
Plasmoid chains develop when the sheet length-to-width aspect ratio $A\gtrsim30$.
In the strongly magnetized limit, the onset of reconnection occurs in
roughly 2--6 light-crossing times, depending on the imposed driving timescale,  which controls the duration of the thinning phase. In the subsequent nonlinear merging phase, the evolution becomes effectively independent of the initially imposed velocity, leading to magnetic-energy dissipation consistent with a normalized reconnection rate $\sim 0.1$.
Our results have important implications for explosive release of magnetic energy in magnetospheres of astrophysical compact objects and their surroundings.
\end{abstract}

\section{Introduction}
\label{sec:intro}
Magnetic reconnection is ubiquitous in astrophysical plasmas. For example, it is believed to power the explosive release of  magnetic energy in solar flares \citep{Parker1957}. By analogy with solar flares, reconnection-powered emission mechanisms have been associated with more distant high-energy signals, such as flares from pulsar-wind nebulae \citep{Uzdensky2011,Cerutti2013}, and high-energy outbursts from neutron-star and black-hole magnetospheres \citep{Thompson1994,Lyutikov2003}. Reconnection in these high-energy environments takes place in the relativistic regime \citep{Sironi2025}. Kinetic particle simulations have shown that reconnection in relativistic astrophysical plasmas can indeed accelerate particles efficiently \citep[e.g.,][]{Guo2014,Sironi2014,Werner2016} 
and power high-energy emission \citep{Cerutti2013, Zhang2020, Chernoglazov2023, Comisso2023, Hakobyan2023b, Mehlhaff2024}, yet much about the physics of relativistic reconnection remains to be understood.

Magnetic reconnection is enabled by nonideal electric fields, which develop most notably near current sheets. 
Resistive magnetohydrodynamic (MHD) studies revealed key aspects of the onset of reconnection in evolving current sheets \citep[e.g.,][]{Biskamp1986,Shibata2001,Loureiro2007, ComissoLHB2016, Uzdensky2016, Huang2017,Rosenberg2021, Leake2024, Tolman2024}, but many astrophysical plasmas are effectively collision-free. Reduced fluid models incorporating Hall effects and electron-MHD physics capture some kinetic behavior \citep{DelSarto2016,Pucci2017,Shi2019,Mallet2020}, yet a detailed fully kinetic investigation of how sheets form and eventually reconnect is presently lacking.

Moreover, most kinetic and fluid simulations begin from an already formed current sheet which reconnects spontaneously. This usually involves setups featuring Harris sheets \citep{Porcelli1991,Schep1994}, coalescing islands \citep{Uzdensky2000,Bhattacharjee2009,Huang2010}, or repelling force-free channels \citep{Keppens2014,Ripperda2017a,Ripperda2017b}, to name a few. However, current sheets often form and reconnect under the action of external forces. 
Examples include the Taylor problem \citep{HK85,Birn05,CGW15} or merging flux tubes pushed together by converging flows \citep{Lyutikov2017,Ripperda2019}.

In the surroundings of compact objects, magnetic structures can coalesce explosively when twisted and sheared by strong flows and relativistic frame-dragging effects. 
Relevant examples include twisted magnetar magnetospheres \citep{Parfrey2013}, colliding flux tubes in pulsar magnetospheres \citep{Mahlmann2023}, frame-dragging–twisted tubes between a black-hole ergosphere and disk \citep{Koide2006}, and loops in a turbulent accretion-disk corona \citep{UzdenskyGoodman2008,Yuan2009}. In these magnetically dominated plasmas, the Alfvén speed approaches the speed of light $c$, which enables rapid variability on light-crossing time scales.
Thus, the problem of collisionless reconnection onset in the strongly magnetized regime is of prime interest for understanding the observed time scales of electromagnetic flares in relativistic astrophysical plasmas (e.g., \citealt{Krawczynski2004, Abdo2011, Yuan2016,GRAVITY2021, Ripperda2022}). Here, we aim to fill a gap in the literature by studying the dynamics of current sheets in collisionless pair plasmas over their full lifetime, from formation to the final stages of the reconnection process, using 2D Particle-in-Cell (PIC) simulations. While motivated by high-energy astrophysical phenomena, the physics of externally driven reconnection and current sheet formation studied here also has potential relevance to laboratory experiments and solar loops interactions \citep[e.g.,][]{Hare2017, Kumar2010}.

Global PIC simulations can self-consistently describe the formation of current sheets in realistic geometries, but they cannot presently achieve realistic astrophysical scale separations. For example, in the largest (to date) 3D simulations of pulsar magnetospheres, the light-cylinder radius is only about $200$ times larger than the plasma skin depth \citep{Hakobyan2023}. Similarly, in global PIC simulations of black-hole magnetospheres,
the gravitational radius typically measures only about a few tens of plasma skin depths in size \citep{Parfrey2019, Crinquand2022, ElMellah2023, Galishnikova2023}. 
Therefore, in order to achieve a maximum possible scale separation and to study the onset problem under well-controlled initial conditions, we focus here on local PIC simulations, featuring two isolated and initially force-free magnetic flux tubes.

The rest of the paper is organized as follows. In Section~\ref{sec:setup} we describe our numerical setup, initial conditions, and simulation parameters. In Section~\ref{sec:results} we present the results of our 2D PIC experiments, including (i) the three‐phase evolution from current‐sheet formation to plasmoid development and fast merging, (ii) quantitative analysis of sheet thinning and aspect‐ratio scaling, (iii) plasmoid statistics, and (iv) measurements of the reconnection and energy‐dissipation rates. In Section~\ref{sec:conclusion} we discuss the astrophysical implications of our findings and summarize our main conclusions.

\section{Methodology}\label{sec:setup}

\begin{figure*}[ht]
    \centering
    \includegraphics[width=1\textwidth]{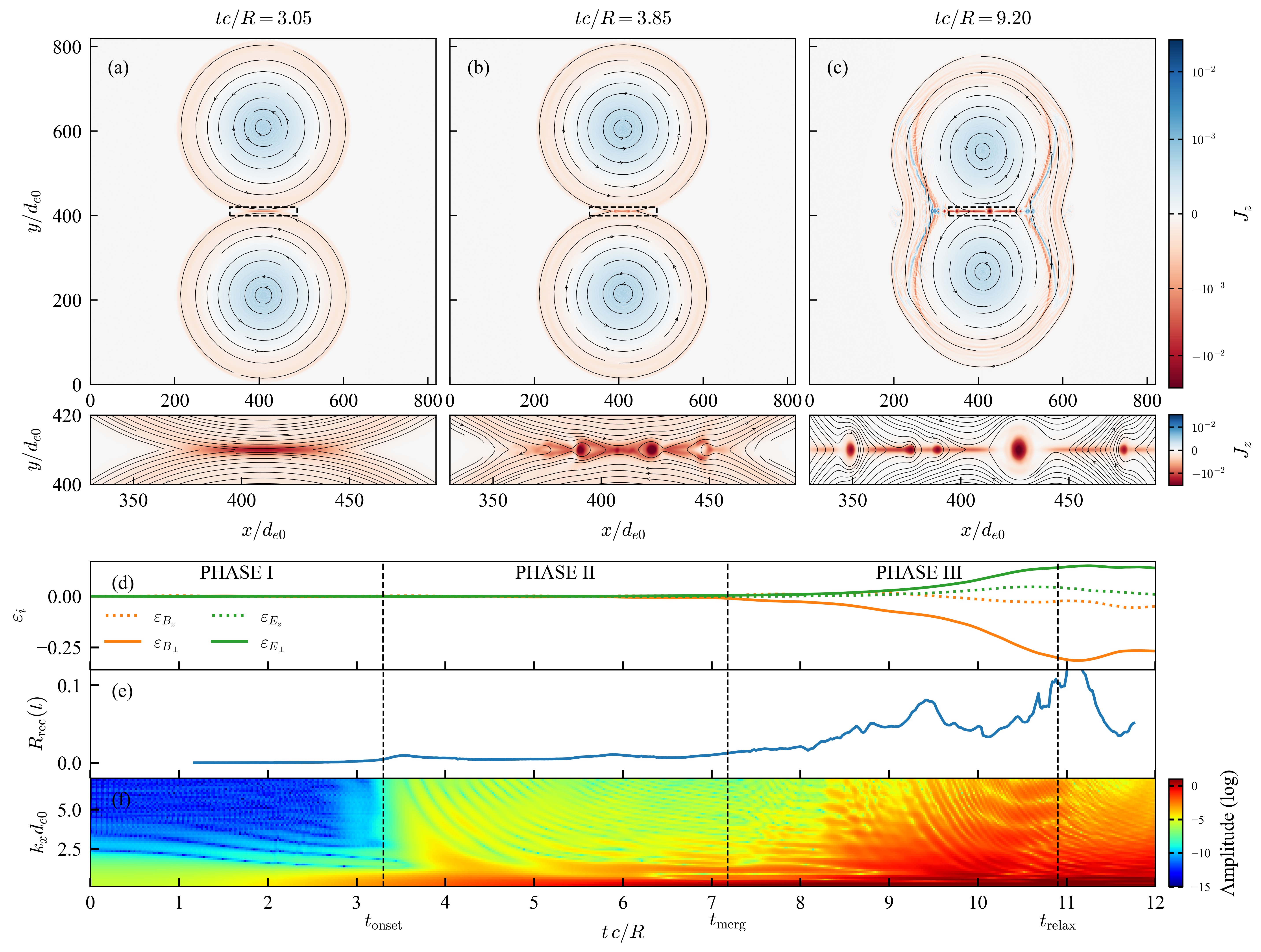}
    \caption{(a)--(c): Contours of out-of-plane current density $J_z$ (color scale) and magnetic field lines (black) during the evolution of the flux tubes. The dashed black boxes mark the regions shown in the zoomed-in view below. (d): Temporal evolution of the energy components, $\varepsilon_i = (E_i(t)-E_i(0))/E_{B_\perp} (0)$. (e): Time evolution of the normalized reconnection rate. (f): $k_x$-spectra of the magnetic flux function averaged inside the current sheet (over $10 \dep$ along $y$). The color-coded intensity represents the logarithmic amplitude of the wavenumbers.  Key time thresholds ($t_{\rm onset}$, $t_{\rm merg}$, and $t_{\rm relax}$) are indicated by vertical dashed lines. This simulation was run with $\vp=0.05c$ ($\td=20R/c$), $\sigo=50$ ($\sigin=8)$, and $R/\dep=205$.}
    \label{fig:tubes_combined}
\end{figure*}

All of our simulations are carried out with the PIC code \textsc{Tristan-MP v2} \citep{tristan}. We perform 2D simulations using an electron--positron pair plasma.
We use a configuration inspired by Lundquist’s force-free cylinders
\citep{Lyutikov2017,Ripperda2019}. 
Inside each flux tube of radius $R$, the poloidal magnetic field is given by
\begin{equation} \label{poloidal}
B_{\phi}(r) = B_0\mathcal{J}_1(\alpha r/R),
\end{equation}
and the vertical (i.e., out-of-plane) magnetic field by 
\begin{equation}\label{toroidal}
B_z(r) = B_0\sqrt{\mathcal{J}_0(\alpha r/R)^2 +C},
\end{equation}
where $r$ is the distance from the center of the tube, $\mathcal{J}_0$ and $\mathcal{J}_1$ are Bessel functions of the the first kind,  $\alpha = 3.8317$ (the first zero of $\mathcal{J}_1$), $C=0.01$, and $B_0$ is a reference magnetic-field strength. The solutions 
\eqref{poloidal} and \eqref{toroidal} are truncated at $r=R$. 
Outside the two flux tubes, the in-plane $(r,\phi)$ magnetic field is zero, and $B_z = B_0 \sqrt{\mathcal{J}_0(\alpha)^2 +C}$. The flux tubes carry no net current at the start of the simulation, so that they do not attract each other. For reference, we show spatial profiles of the initial magnetic-field configuration in Appendix~\ref{app:profiles}. This shows that the guide field ($B_z$) exhibits a nontrivial profile, which is needed to impose the force-free property of the initial condition. The guide field peaks at the centers of the tubes, decreases toward the edges, and exhibits a slight bump near the region of contact between the tubes.
We employ a square simulation domain of size $L_x = L_y = 4R$. The two flux tubes are initially centered at ${\bf r}_1 = (x_1, y_1) = (2R, R)$ and ${\bf r}_2 = (x_2, y_2) = (2R, 3R)$.
To initiate the current-sheet formation, we push the tubes toward each other with a prescribed initial velocity $\vp$, corresponding to a driving timescale $t_{\rm drive} = R/\vp$. The initial electric field inside each flux tube is given by $\vecE=-\bb{v}_\mathrm{push}\times\vecB/c$. We consider a range of different push velocities, from a few 
percent to a few tens of percent of $c$. Such driving speeds are in line with Alfv\' enic-like motions in 
high-$\sigma$ relativistic plasmas \citep[e.g.,][]{Parfrey2013,Mahlmann2023}.

Our simulation box is initially filled with a cold pair plasma of uniform density $n_0$ and
temperature $T_0 = 0.005 m_{e}c^2$. The electron and positron distributions inside each
tube are initialized with bulk motion profiles consistent with Ampère's law and the $\vecE\times\vecB$ drift motion. 
We perform simulations for a range of different plasma
magnetizations $\sigo = B_0^2 / (4\pi n_0 m_{e} c^2)$, box sizes, and initial push velocities.
The magnetization $\sigo$ based on the reference field strength $B_0$
ranges from $25$ to $800$. 
We note that the \emph{average in-plane} 
magnetization \emph{inside} the tubes $\sigin \approx 0.16\sigo$. 
We explore a range of computational domain sizes, from $L_x=204.8\,\dep$ to $L_x=6553.6\,\dep$, where $\dep =c/\omega_{p}$
is the cold plasma skin depth, and $\omega_{p}=\sqrt{4\pi n_0 q^2/m_{e}}$ is the plasma frequency. 
Thus, the flux-tube radius $R$ ranges from roughly $38\,\dep$ to $1638\,\dep$
in our set of runs.
Our numerical resolution is $dx=\dep/5$, where $d x$ is the grid cell size.

Significant computing resources are required in order to explore the flux tube dynamics in relatively 
large-scale domains. For instance, the computational
grid size in our largest run is 32,768$^2$.
For the dependence on numerical resolution and number of particles per cell, see Appendix~\ref{app:convergence}.

\section{Results}\label{sec:results}

\subsection{Description of the Evolution}

We identify three distinct phases in the evolution of the system (see Fig.~\ref{fig:tubes_combined}):
\begin{itemize}
  \item \textit{Phase I (Sheet Formation):} The flux tubes are externally driven to approach each other, resulting in the formation of a thin current sheet (Fig.~\ref{fig:tubes_combined}(a)). As the sheet thins rapidly, small-scale perturbations in the form of tearing modes \citep[see e.g.][]{Furth1963, Coppi1976, Zelenyi1979, Tolman2018, Granier2021, Demidov2024} start growing (Fig.~\ref{fig:tubes_combined}(f) and Fig.~\ref{fig:lineargr} in Appendix~\ref{app:lineargr}). 
  \item \textit{Phase II (Plasmoid Formation):}  When the tearing-driven fluctuations grow to sufficiently large amplitude, the instability transitions to the nonlinear regime and magnetic islands (plasmoids) become clearly visible.  We denote the onset of this transition by the time $\tonset$, which we identify as the time at which the sheet thickness attains its first minimum, 
  before it widens again as the islands grow in size.
  The sheet then fragments into multiple plasmoids (panel (b)), leading to an increase in the reconnection rate (panel (e)). 
  \item \textit{Phase III (Merging):} In the final phase, the flux tubes coalesce rapidly, causing a fast depletion of magnetic energy (panel (d)). We denote the beginning of the merging process by $t_{\rm merg}$ (i.e., when the conversion of in-plane magnetic energy reaches $1\%$), and the time when the flux tubes relax toward a quasi-stable configuration by $t_{\rm relax}$ (i.e., when the conversion reaches 30$\%$). The duration of the merging phase is given by $\Delta t_{\rm merg}=t_{\rm relax}-t_{\rm merg}$. 
\end{itemize}

\subsection{Current-Sheet Formation and Reconnection Onset}

\begin{figure}
    \includegraphics[width=\linewidth]{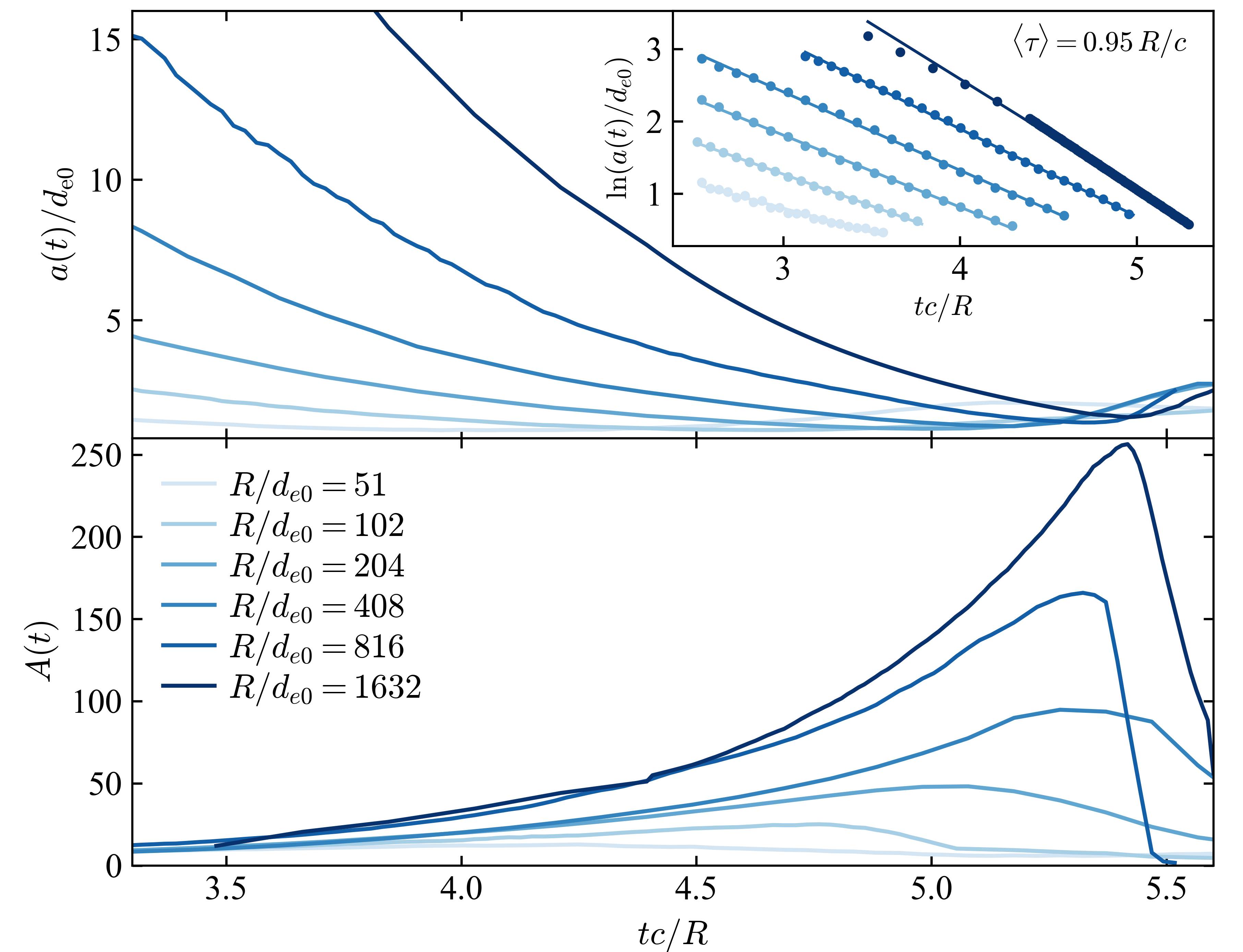}\\
    \caption{
    Time evolution of the normalized current sheet width, $a(t)/d_{e0}$, and the aspect ratio, $A(t)=\ell(t)/a(t)$, for different system sizes at $\sigin=8$. Inset: the evolution of $\ln\left(a(t)/d_{e0}\right)$, showing the exponential thinning.}
    \label{fig:combined_sigma}
\end{figure}
We first examine the formation of the current sheet. The sheet thickness $a(t)$, measured in units of $d_{e0}$, is determined by applying a polynomial fit to the transverse profile of the out-of-plane current density $J_z$ \citep[e.g., see][]{Huang2017}.
The sheet boundaries are at the positions where $J_z$ reaches two-thirds of its maximum value. Fig.~\ref{fig:combined_sigma} shows the evolution in time of the sheet thickness and aspect ratio, $A(t)=\ell(t)/a(t)$, where $\ell$ is the current-sheet length, for several $R/\dep$.

To gain analytical insight into the driven current‐sheet formation, we adopt a 1D symmetric approximation on ideal scales exceeding the plasma microscales, so that the magnetic field is frozen‐in. Given that our main focus is the magnetically dominated regime, we neglect the kinetic pressure in the following analysis. We consider the region between the flux tubes 
and assume that the reconnecting magnetic field reverses linearly across a region of half-width $a(t)$, such that $B_x\simeq\bup y /a(t)$ for $|y|\le a(t),$ and $B_x\simeq \pm \bup$ for $|y|>a(t)$. For our setup, $a(t=0)=a_0 \simeq R/2$ (see equilibrium profiles in Appendix~\ref{app:profiles}). 

During the initial kinematic phase, 
the current-sustaining motions remain negligible compared to the imposed inflow. Thus, the inflow
$v_y(\pm a) = \mp v_{\rm push}$ drives the sheet edges inward at constant speed. Hence, $a(t) \simeq a_0 - v_{\rm push}\,t$
during the kinematic stage, which lasts until time $t_{\rm k}$.
The transition time $t_{\rm k}$ is defined as the moment when the Lorentz‐pinch acceleration becomes comparable to the driving compression, and the sheet half‐width at that moment is $a_{\rm k} \simeq a_0 - v_{\rm push}\,t_{\rm k}$.

At $t\ge t_{\rm k}$, the sheet has been sufficiently compressed so that the drift current $J_z\sim B_{x\rm up} /(4 \pi a_K)$ generates a Lorentz‐pinch force $(\bb{J}\times \vecB)_y=J_z\,B_{x\rm up} = \rho\,v_y\partial_y v_y$ dominating the external push, causing self‐collapse. Since the guide field within the current sheet remains very weak and is continually advected out of the layer by the outflow, the term $J_x B_z$ is negligible compared to the $J_x B_{x\rm up}$ pinch force.
Using $B_x\simeq B_{x\rm up} y/a$ and $J_z\simeq (c/4\pi)(B_{x\rm up}/a)$, the 1D fluid momentum equation along $y$ gives $ \partial v_y= (c^2/w)\left(J_z\,B_{x \rm up} \right)\simeq v_{\rm A}^2/a^2 y$, where $v_{\rm A}^2 = B_{x\mathrm{up}}^2 c^2/(4\pi w + B^2)$,   and $w$ is the relativistic enthalpy density. For simplicity, we assumed sub-relativistic motion of the converging flow, so that the bulk Lorentz factor is of order unity.
We assume a self‐similar inflow profile of the form $v_y(y)\simeq -y/\tau$. Hence, the sheet edge obeys $\rmd a/\rmd t = v_y\bigl(a(t)\bigr) = -a(t)/ \tau$, which gives
\begin{equation}
a(t-t_{\rm k}) = a_{\rm k}\,\exp\bigl[-(t - t_{\rm k})/\tau\bigr], \quad
\text{with}\quad \tau = \frac{a_{\rm k}}{v_{\rm A}}.
\end{equation}
In the inset of Fig.~\ref{fig:combined_sigma} we show $\ln\bigl(a(t)/d_{e0}\bigr)$ during the accelerated stage of  the collapse, confirming the exponential thinning.

To determine quantitatively when the kinematic thinning gives way to the exponential collapse, we compare the time it takes the external drive to compress the sheet from its initial half‐width $a_0$ down to $a_{\rm k}$, namely $t_{\rm push} \simeq(a_0-a_{\rm k})/v_{\rm push}$, with the Alfvén‐crossing time across the compressed half‐width, $\sim a_{\rm k}/v_{\rm A}$. At the transition, these two timescales are equal. This gives
\begin{equation}
a_{\rm k} \simeq \frac{a_0 v_{\rm A}}{v_{\rm push} + v_{\rm A}},
\quad
t_{\rm k} \simeq \frac{a_0}{v_{\rm push} + v_{\rm A}}.
\end{equation}

In the limit $v_{\rm push}\ll v_{\rm A}$, the pinch wave outruns the drive and the sheet undergoes kinematic thinning for roughly one Alfv\' en‐crossing time of the original width, $t_{\rm k}\simeq R/v_{\rm A}$. When $v_{\rm push}\sim v_{\rm A}$, the drive and the pinch forces balance, yielding a transition approximately halfway through the compression, $t_{\rm k}\simeq R/(2\,v_{\rm A})$. In this work, we specifically probe in details these two regimes. 
Conversely, if $v_{\rm push}\gg v_{\rm A}$, the drive dominates until the flux tubes collide at super-magnetosonic speed, launching a fast‐mode MHD shock that disrupts the sheet, produces sharp, wavelike structures in the current density that propagate toward the flux‐tube cores (see Appendix~\ref{app:shock}, Fig.~\ref{fig:jslices}), and heats the plasma.

We now estimate the onset condition by tracking the amplitude $W(t)$ of the linear tearing perturbation during the exponential collapse of the current sheet.  Once the sheet half‐thickness enters the self‐similar regime, the instantaneous growth rate of $W$ is given by $\rmd_t(\ln W) = \gamma\bigl(a(t)\bigr) - 1/\tau.$ We adopt the scaling $\gamma(a)\approx v_{\rm th} (d_e^2/a^3)(B_{x\rm up}/B_z)$ which corresponds to the fastest‐growing wavenumber $k_{\max}\sim1/(2a)$ (\cite{Liu2013, Demidov2024}). To improve readability, we define  $G\equiv v_{th} d_e^2 B_{x\rm up}/B_z $. Here we neglect any variation in the ratio $B_{x\rm up}/B_z$, and assume that tearing begins precisely at $t=t_K$ when exponential thinning starts. The sheet is considered disrupted once $W$ has grown by a factor $\exp(N_e)$ \citep{ComissoLHB2016,Uzdensky2016, Tolman2018}.  Integrating from $t_{\rm k}$ to $t_{\rm onset}$,
\begin{equation}
\int_{t_{\rm k}}^{t_{\rm onset}}\bigl[\gamma(a(t))-\tau^{-1}\bigr] \rmd t = N_e,
\end{equation}
and using $\exp[3(t_{\rm onset}-t_{\rm k})/\tau]\gg1$, one finds
\begin{equation}
a_{\rm onset} = \Bigl(\frac{G \tau}{3 N_e}\Bigr)^{1/3}, \quad t_{\rm onset} = t_{\rm k} + \frac{\tau}{3}\ln\Bigl(\frac{3 a_{\rm k}^3 N_e}{G \tau}\Bigr).
\end{equation}
For fiducial values $B_{x\rm up}/B_0=0.7$, $B_z/B_0=0.2,$ $T=0.005 mc^2,$ $v_{\rm push}=0.02c$,  $N_e=1$, and $R/d_e=408$, we obtain $t_{\rm onset}\approx5.7 R/c$. For comparison, if tearing is ignored and the onset is defined instead by the condition $a(t_{\rm onset})=d_e$, then $t_{\rm onset} = t_{\rm k} + \tau\ln(a_{\rm k}/d_e),$ which yields $t_{\rm onset}\approx 6.1 R/c$.
The difference of $\sim0.5 R/c$ implies that linear tearing has limited influence on the onset time of a rapidly thinning current sheet.

\begin{figure}
    \centering
    \includegraphics[width=\linewidth]{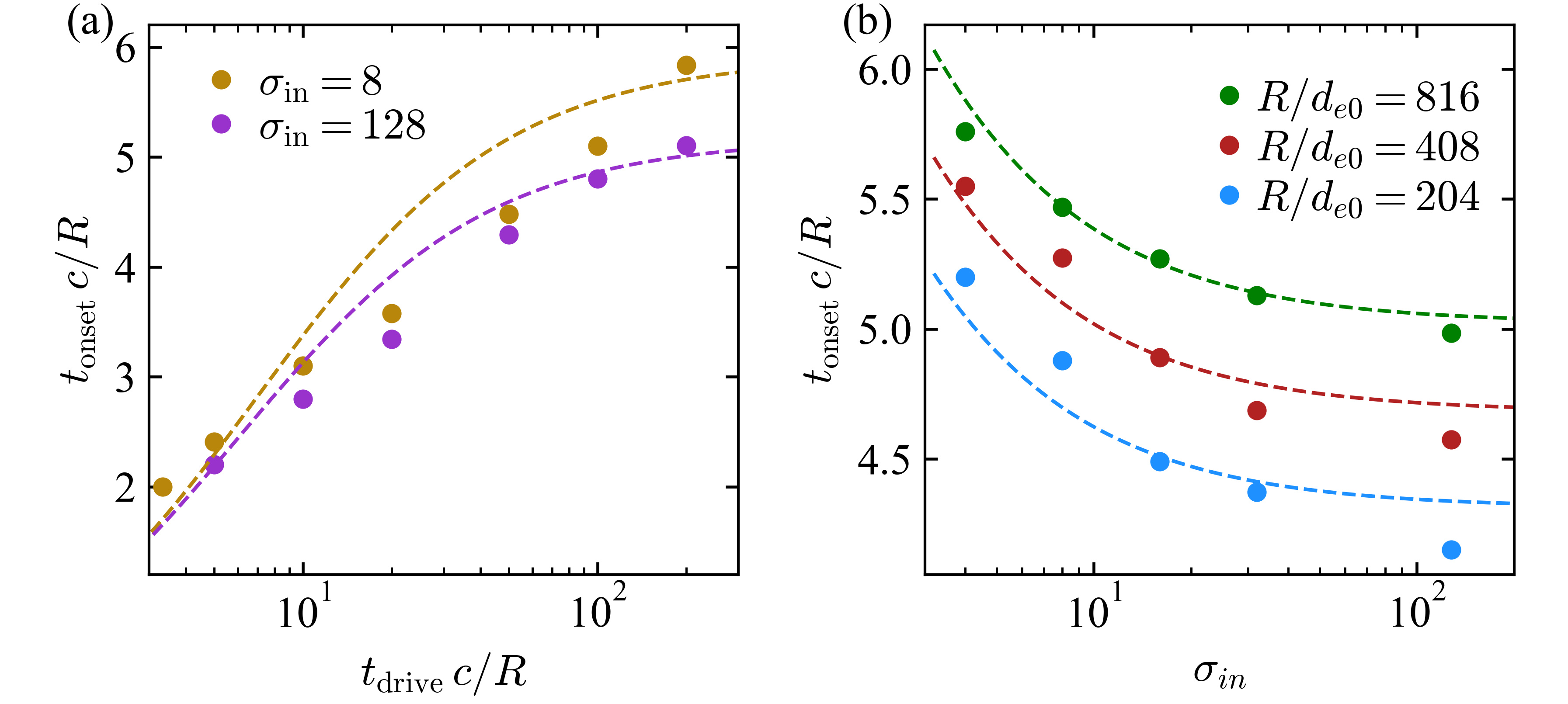}\\[1ex]
    \caption{(a): Reconnection onset times as a function of the driving timescale for $\sigin=8$ and $\sigin=128$ at $R/\dep = 204$. (b): Onset times as a function of in-plane magnetization for $t_{\rm drive} c/R = 50$. The dashed lines show the theoretical estimate, Eq.~\eqref{tons}.}
    \label{fig:collapse_time}
\end{figure}

For our strongly magnetized, low-guide-field, cold, and rapidly forming current sheet, we approximate $t_{\rm onset}$ by 
\begin{equation} \label{tons}
t_{\rm onset}
\simeq \frac{a_0}{v_{\rm A} + v_{\rm push}}\Bigl[1 + \ln\!\frac{a_{\rm k}}{d_{e0}}\Bigr],
\end{equation} 
\newline
where the first term is the duration of the kinematic linear thinning phase and the second term is the duration of the exponential collapse phase.  
Using $v_{\rm A} = c\sqrt{\sigma_{\rm in}/(1+\sigma_{\rm in})}$, we see that increasing the in‐plane magnetization leads to a faster exponential collapse. In the high‐$\sigma_{\rm in}$ limit, $v_{\rm A}\approx c$, so the thinning time scale $\tau\sim a_0/(c+v_{\rm push})$; for low $\sigma_{\rm in}$, $v_{\rm A}\approx c\sqrt{\sigma_{\rm in}}$ yields $a_K\propto a_0\sqrt{\sigma_{\rm in}}$ and slower thinning. In the pinch‐dominated regime ($v_{\rm push}\ll v_{\rm A}$), $t_{\rm onset}\sim (a_0/v_{\rm A})\ln(a_0/d_{e0})$; and in the strongly magnetized limit ($\sigma_{\rm in}\gg1$), $t_{\rm onset}\sim R/c\,$. As seen in Fig.~\ref{fig:combined_sigma}, the rate of exponential collapse in units of $R/c$  is slightly faster in larger-size systems.
Fig.~\ref{fig:collapse_time}(a) shows that $t_{\rm onset}$ increases with the driving timescale $t_{\rm drive}=R/\vp$, in good agreement with Eq.~\eqref{tons}. Panel (b) shows that $t_{\rm onset}$ depends only weakly on the exact value of the 
magnetization when $\sigin\gtrsim 1$; varying $\sigin$ by more than one order of magnitude changes $t_{\rm onset}$ by only tens of percent. This arises because the magnetization enters $t_{\rm onset}$ in Eq.~\eqref{tons} only via the Alfv\' en speed, which is always close to $c$ for our set of runs.
We also find that the normalized collapse time, $t_{\rm onset}\,c/R$, grows only logarithmically with the system size $R/\dep$,
which is consistent with the second term in Eq.~\eqref{tons}.

For system sizes from $R/\dep\approx50$ to $ \approx 1600$, we find that the sheet thickness $a(\tonset) = a_*$ 
at reconnection onset depends very weakly on the system size; it scales roughly as $a_* \sim \dep (R/\dep)^{0.08}$  (see also Fig.~\ref{fig:combined_sigma}).
Our simulations show that the dependence of $a_*$ on $\sigin$ is also minimal. Moreover, when varying the drive velocity from $v_{\rm push}=0.02\,c$ to $0.2 c$, the measured collapse width changes only from $1.10 \dep$ to $1.16 \dep$, demonstrating that $a_*$ is also essentially independent of the push magnitude.

We also measure the length $\ell$ of the evolving current sheet. The length at collapse $\ell_*$ increases with the system size roughly as $\ell_* \sim 0.3 (R/\dep)^{1.08}$. Therefore, the critical sheet aspect ratio $A_* = \ell_*/a_*$  scales roughly  as $A_*\sim 0.3 R/\dep$. Based on earlier resistive works (\citealt{Biskamp1986}), a Sweet--Parker current sheet becomes unstable to plasmoid formation once its aspect ratio  $\ell/a_{SP}\sim100$. In our simulations, this threshold appears to be roughly at $A_* \gtrsim 30$. 

For the cases shown in Fig.~\ref{fig:combined_sigma} with $\vp = 0.02c$, the simulation with $R/\dep = 51$ does not form plasmoids, but plasmoids appear in simulations with larger boxes. As we discuss in the next section, the push velocity influences the number of plasmoids formed, since it affects the amplitude of perturbations and to some extent also the geometry of the current sheet.
For the system size $R/\dep = 204$, at the lowest background speed $\vp = 0.005c$, the current-sheet length at disruption is approximately $\ell_* \approx 40  \dep$, while at the highest speed $\vp = 0.2c$, we find $\ell_* \approx 96 \dep$. 

Therefore, shorter driving time scales, corresponding to higher background flow speeds, result in somewhat larger sheet aspect ratios.

\subsection{Plasmoids}

\begin{figure}[ht]
    \centering
    \includegraphics[width=0.5\textwidth]{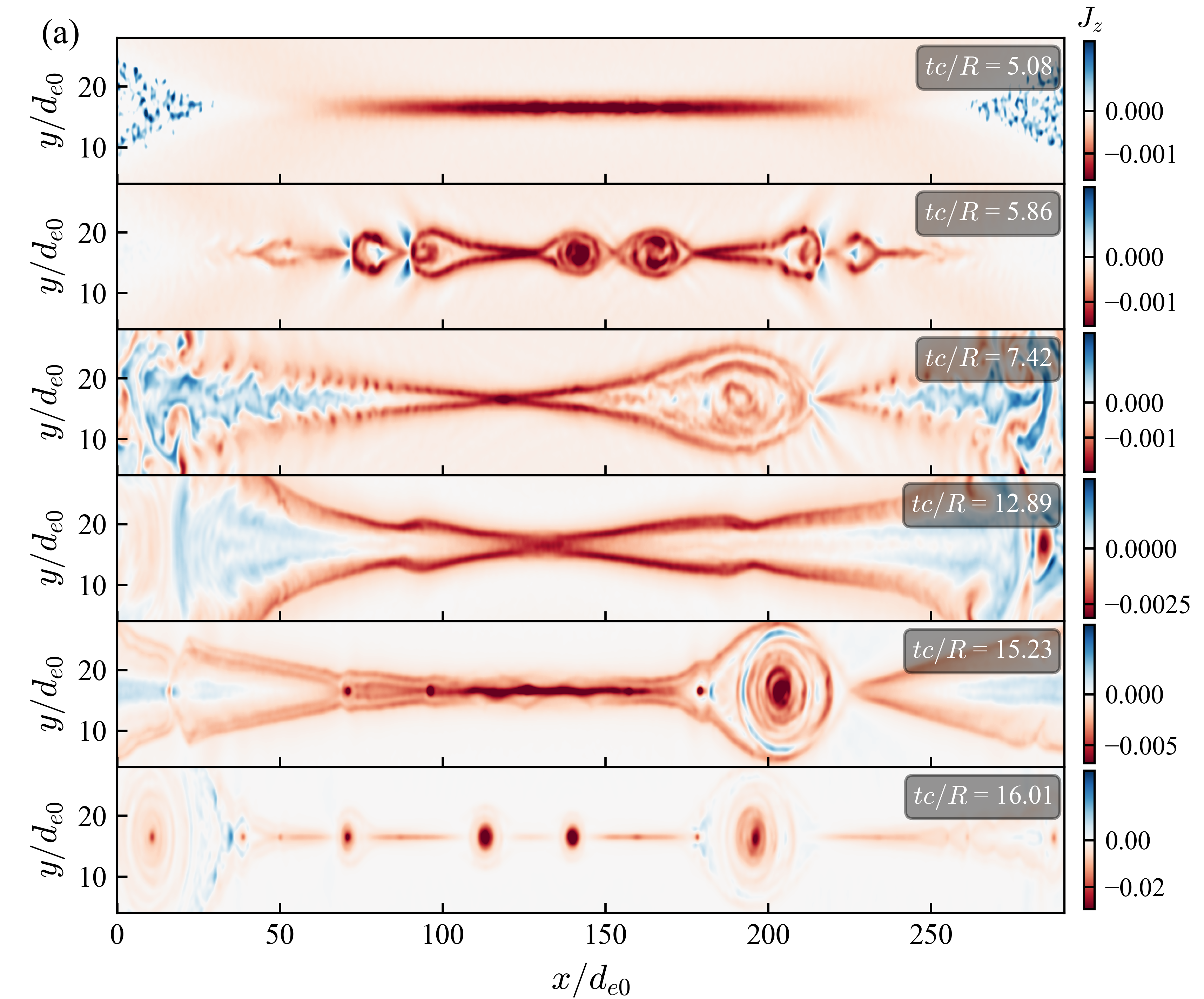}\\
    \includegraphics[width=0.5\textwidth]{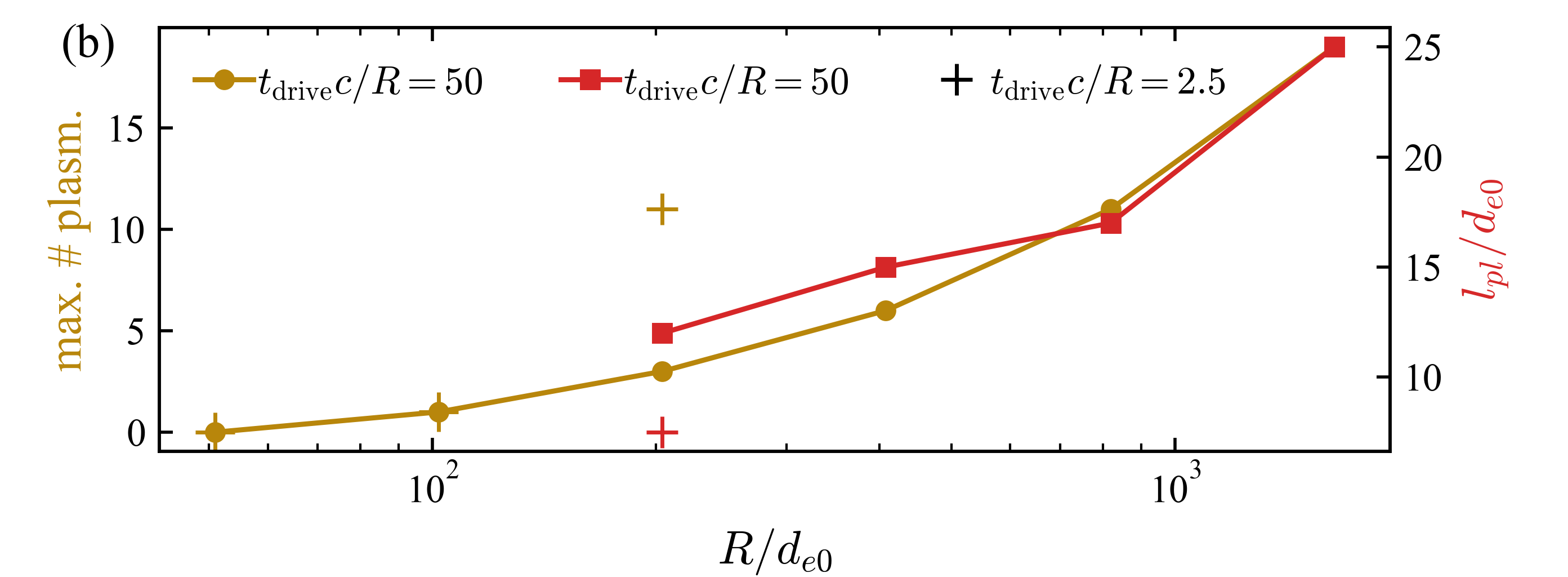}\\[1ex]
    \caption{(a): Evolution of the out=-of-plane current density $J_z$ for $R/\dep=408$ and $\vp=0.02c$ (data shown have been downsampled by a factor of 8). (b): Maximum number of plasmoids and the minimum interplasmoid distance at the start of Phase II as functions of the system size $R/\dep$, for $\td=50 R/c$ ($\vp=0.02c$) and $\sigin=8$. Crosses indicate the same quantities for a shorter driving time, $\td=2.5 R/c$ ($\vp=0.4c$).}
    \label{fig:nbrplas}
\end{figure}

In panel (f) of Fig.~\ref{fig:tubes_combined}, the $k_x$-spectrum of the magnetic flux function, averaged over $10\,\dep$ across the current sheet, is shown. We observe that during Phase I the modes satisfying $k_x\,\dep \lesssim 1$ exhibit linear growth. In 
Fig.~\ref{fig:lineargr} of Appendix~\ref{app:lineargr}, we show how the growth rates of modes with different $k_x$
evolve in a typical simulations, confirming that the disruption is preceded by a phase of linear growth. 
At time $t_{\rm onset}$, the nonlinear Phase II begins with numerous unstable modes, leading to the formation of plasmoids (see Fig.~\ref{fig:nbrplas}(a) at $t c/R = 5.86$). These plasmoids are then expelled from the current sheet by the outflow faster than they can grow. Once expelled, the current aligns along the separatrices, producing a characteristic open outflow structure ($t c/R = 7.42$--$12.89$). When fast merging starts, the current sheet stretches and thins again, reaching a minimum thickness of approximately $0.6\dep$. This brings higher–$k_x$ modes into the nonlinear regime, leading to a new generation of plasmoids ($t c/R = 15.23$). Overall, plasmoid formation is closely linked to whether we see modes with $k_x \dep \gtrsim 1$ reaching large amplitudes during Phases II and III.

In Fig.~\ref{fig:nbrplas}(b), the red curve shows the maximum number of plasmoids observed simultaneously during Phase II, while the blue curve shows their spacing, estimated from the separation between magnetic O‑points. The circles denote slow driving ($t_{\rm drive}=50 R/c$) and the crosses denote fast driving ($t_{\rm drive}=2.5 R/c$).  For $R/d_{e0}=204$, fast driving yields four times as many plasmoids with a shorter plasmoid separation of about $l_{pl} \approx 7\dep$, suggesting that a stronger external push leads to a denser plasmoid chain. We observe that the number of plasmoids during Phase II scales approximately as $N \propto R/\dep \propto A$. 
The effective aspect ratio of the interplasmoid current sheets is roughly $A \lesssim 10$,
in agreement with previous studies \citep[e.g.,][]{Petropoulou2018,Granier2022}.

\subsection{Reconnection Rate}

\begin{figure}[ht]
    \centering
    \includegraphics[width=0.5\textwidth]{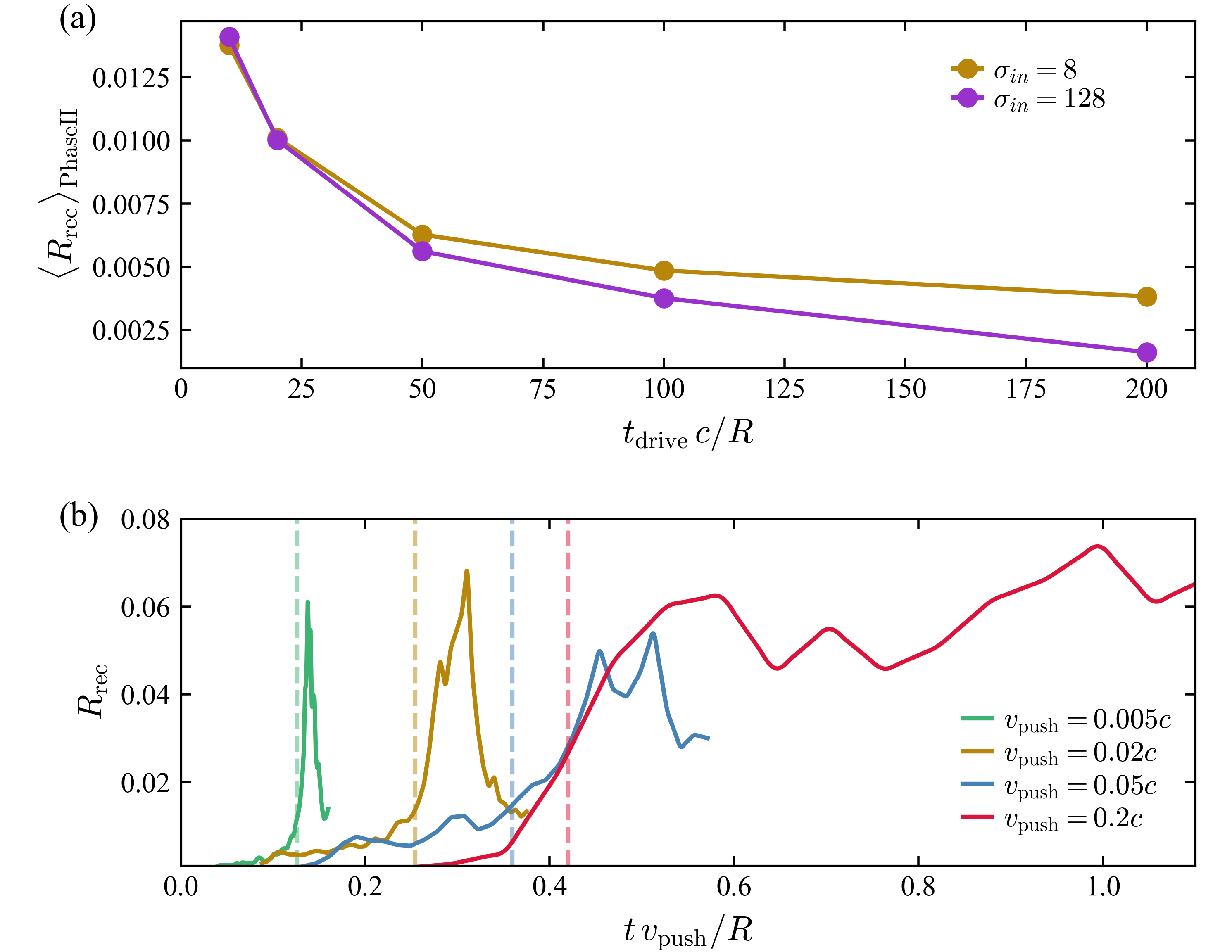}\\
    \caption{Reconnection rate $R_{\rm rec }$ as measured via Eq.~\eqref{recmeassure}. (a): values averaged over time during Phase II
    (i.e., from $\tonset$ to $t_{\rm merg}$, where 
     $t_{\rm merg}$ marks the beginning of the merger Phase III). (b): evolution of $R_{\rm rec}$ in time  for various $v_{\rm push}$. Dashed vertical lines mark $t_{\rm merg}$.}
    \label{fig:rec_combined}
\end{figure}
The normalized reconnection rate in our runs is measured as
\begin{equation}\label{recmeassure}
R_{\rm rec}(t) = \frac{1}{v_{\rm A}\,\bup}\frac{\rmd}{\rmd t}\Big[\max(A_z)-\min(A_z)\Big],
\end{equation}
where $A_z$ is the out-of-plane component of the magnetic vector potential in the midplane between the two tubes (i.e., along the current sheet), and $\bup$  and $v_{\rm A}$ are the
upstream reconnecting field and Alfv\'en speed (based on $\bup$). Eq.~\eqref{recmeassure} measures how fast the magnetic flux is being transferred across the separatrix. The time-averaged value of $R_{\rm rec }$ during Phase II for different push velocities at $R/d_{e0}=204$ is shown in panel (a) of Fig.~\ref{fig:rec_combined}. We find that the reconnection rate during Phase II is strongly sensitive to the push velocity $\vp$, while its dependence on magnetization and system size is weak. As $\sigin$ increases from 8 to 128, the rate decreases slightly from $\sim 0.036$ to $\sim 0.020$. Increasing $R$ by a factor of 2 yields a minor drop from $\sim 0.024$ to $\sim 0.020$. 
Panel (b) of Fig.~\ref{fig:rec_combined} shows the time evolution of the reconnection rate for different push velocities. 
The measured reconnection rates are somewhat below the typical $\sim 0.1$ value (with $v_{\rm A}$ based on the in-plane field), which is consistent with the presence of a moderate  guide field in our numerical experiments \citep[e.g., see][as well as Appendix~\ref{app:profiles} for the initial guide-field profile]{Sironi2025}.

\subsection{Fast Merging}

\begin{figure}
    \centering
    \includegraphics[width=0.5\textwidth]{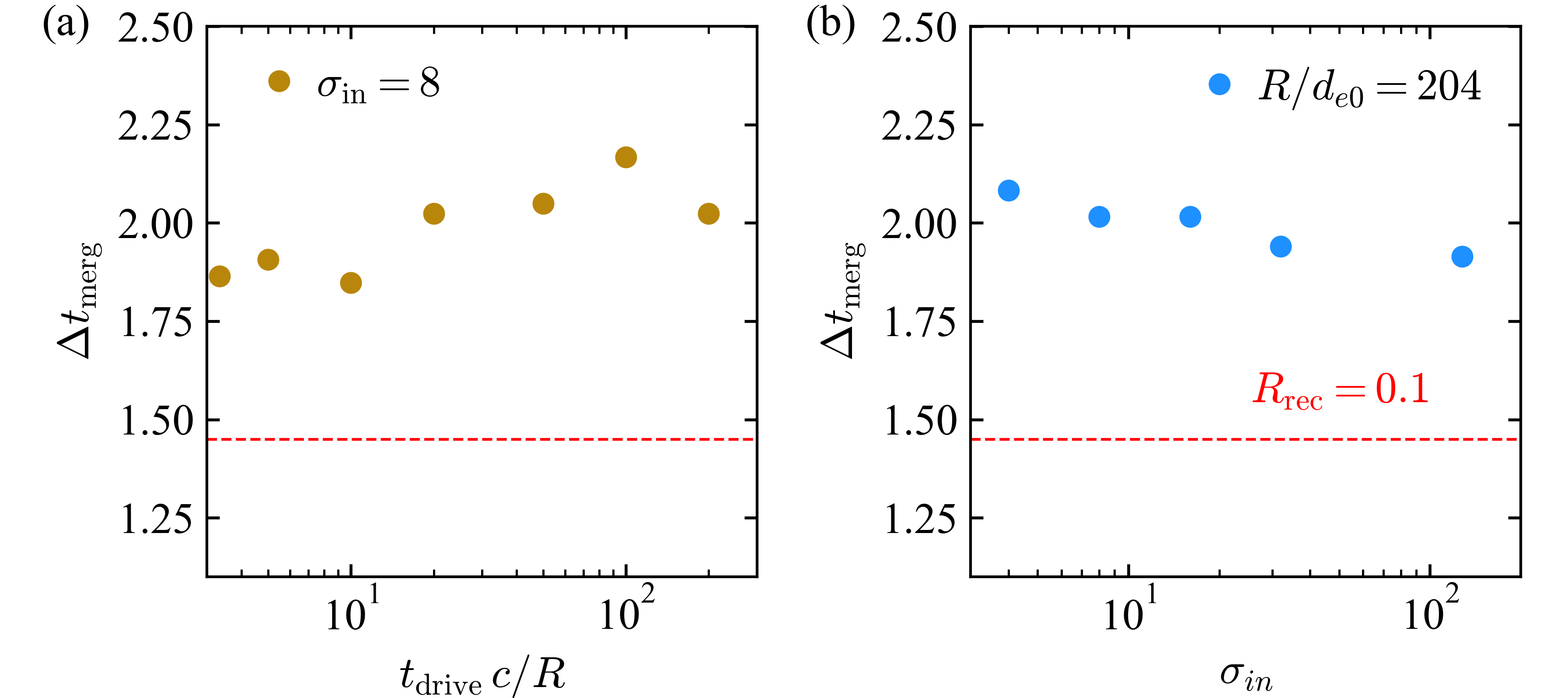}\\[1ex]
    \caption{(a): Normalized merging duration $\Delta t_{\rm merg}$ 
    in units of $R/c$ as a function of the driving timescale for $\sigin=8$ and $R/\dep = 204$. (b): $\Delta t_{\rm merg}$ as a function of in-plane magnetization for $t_{\rm drive} c/R = 50$. The start of the merger phase
    is defined as the time at which 1\% of the initial magnetic energy has been
    dissipated. The end of the merger phase is determined as the time at which the amount of dissipated energy 
    saturates at about 30\% (see Fig.~\ref{fig:tubes_combined}).}
    \label{fig:collapse_time2}
\end{figure}

All simulations, regardless of magnetization, system size, or push strength, converge to a  similar reconnection rate during the final fast-merging phase. The typical value  of the rate is close to, but somewhat below, $0.1$, even for strong driving with $\vp > 0.1c$. The near-universal reconnection rate  during Phase III implies that the available magnetic energy (inside the tubes)  is dissipated at a rate that is largely independent of the initial condition.
To capture the net rate at which the in-plane magnetic energy is dissipated during the merging, we define
\begin{equation}
\mathcal{R}_{\rm diss} = \frac{R}{c E_{B_\perp} (0)}\frac{{\rm d}(-E_{B_\perp} (t))}{{\rm d}t}
\end{equation}
which provides an alternative estimate for the effective reconnection rate based on the net loss of reconnecting magnetic energy. While Eq.~\eqref{recmeassure} measures the instantaneous reconnection rate at the main X-point, $\mathcal{R}_{\rm  diss}\simeq 2R_{\rm rec}$ measures the energy-conversion efficiency over the merging 
process\footnote{The factor $2$ comes from the Poynting flux bringing reconnecting magnetic energy from both sides of the current sheet at about the same rate.}. 
As shown in Fig.~\ref{fig:collapse_time2}, the duration of the merger phase $\Delta t_{\rm merg}$ 
is roughly $2R/c$, during which about 30\% of the
initial magnetic energy is dissipated (see Fig.~\ref{fig:tubes_combined}). Thus, the energy is
dissipated at an average 
rate of $\langle R_{\rm diss}\rangle\approx 0.15$.

We now explain the physical origin of the merging phase’s universality. During Phase II, reconnection progressively removes the poloidal magnetic field $B_\phi$ from the outer layers of the flux tubes. This field contributes two forces: magnetic pressure, $p_{B_\phi} = B_\phi^2/(8\pi)$, which pushes outward, and magnetic tension, with a force density $f_{\rm tens} = (1/4\pi) B_\phi \, \partial_r B_\phi$, which pulls inward along the radial direction. As $B_\phi$ is rearranged, outward pressure support weakens, while inward tension persists, leading to a net radial force toward the center.
This imbalance strengthens over time and eventually initiates a rapid collapse once it overcomes plasma inertia.
The collapse triggers a feedback loop: increased inflow enhances reconnection, further accelerating the collapse.

The guide field $B_z$, although not reconnected, is advected by the inflowing plasma. Its associated magnetic pressure, $p_{B_z} = B_z^2/(8\pi)$, also acts radially outward and can slow down the collapse dynamic. However, $B_z$ does not contribute tension in the radial direction and therefore can delay, but not prevent, the onset of collapse.  

We can express the condition for the collapse onset as an inequality comparing the total inward magnetic tension force to the plasma's resistance to being accelerated:
\begin{equation}
\int_{r_f(t)}^R \frac{B_\phi}{4\pi} \frac{\rmd B_\phi}{\rmd r} \, \rmd r \gtrsim \rho \, a_{\rm crit},
\end{equation}
where $r_f(t)$ denotes the radius down to which the poloidal field has been reconnected, and $a_{\rm crit}$ is the critical acceleration needed to launch the inflow across a distance $\sim R$ over a time $\sim R/v_A$. This gives $a_{\rm crit} \sim v_A^2 / R$, where $v_A$ is the Alfvén speed. Using the explicit profiles for $B_\phi(r)$, we find that collapse begins once the front reaches a critical radius $r_{f,\rm crit} \approx 0.65$--$0.90,R$, where the net inward tension becomes comparable to inertial resistance.

The consistent duration and rate acceleration across all system sizes and $\vp$  suggest that this transition is governed primarily by the internal magnetic structure (the radial profiles of $B_\phi$ and $B_z$) rather than by external forcing. Therefore, any fixed geometry should set a universal scale for $\Delta t_{\rm merg}/(R/c)$, independent of $\sigma_0$ or $R$.

\section{Discussion and Conclusion}\label{sec:conclusion}

We investigated the onset of driven collisionless reconnection in a strongly magnetized pair plasma.
Our 2D PIC simulations feature two force-free flux tubes of radius $R$, which are initially  pushed together with a prescribed velocity $v_{\rm push}$ to trigger current-sheet formation.
We study the full lifespan of the evolving system, from current-sheet formation,  to reconnection onset, and finally into the nonlinear flux-tube merger phase.
At high magnetizations ($\sigin\gg 1$), the reconnection onset time $\tonset$ is of the  order of the light-crossing time $R/c$. In units of light-crossing time, $\tonset$ depends weakly on the flux-tube radius and on 
the magnetization, but changes with the imposed  driving timescale $\td = R/\vp$ as $\tonset \sim R/(v_{\rm A} + \vp)$. This indicates that the external driving significantly influences the duration of the current-sheet formation, as well as the duration of the linear tearing phase.
In realistic astrophysical environments, the driving timescale may be determined by various processes. 
In magnetospheres of compact objects and their surroundings, this involves most
notably the formation of current sheets via 
shearing and twisting of magnetic-field lines, induced by the 
rapid rotation of the central object \citep{Parfrey2013, Parfrey2015, Yuan2019, Mahlmann2020, ripperda2020, Ripperda2022, ElMellah2023}.

Even in our largest simulations with $R/\dep\approx 1600$, we find that the forming current sheet %
thins exponentially until its thickness $a$ reaches the skin-depth scale: $a_*\sim d_{e0}$. 
The aspect ratio at the moment of 
disruption scales as $A_*=\ell_*/a_*\sim 0.3 R/d_{e0}$. When $A_*\gtrsim30$, plasmoids appear, whereas below this threshold only a single X‐point forms. This implies that the smallest reconnecting current sheet that becomes plasmoid-unstable in our simulations has 
a length of at least $30\,d_{e0}$.
For asymptotically large domains, we expect that disruption occurs when the sheet thickness
$a_* \gg \dep$, as anticipated theoretically \citep[e.g.,][]{Comisso2019}. Our results thus indicate that the asymptotic regime 
in a magnetically dominated pair plasma requires $R \gg 10^3 \dep$; the exact size for the presumed transition into the
asymptotic regime is presently unknown (i.e., our results
only provide a lower bound). Further work is needed to clarify how our results extrapolate to much larger domains.

We measure reconnection rates which are largely insensitive to system size and magnetization, but scale with the driving velocity during the 
initial (premerger) stage of reconnection.
In the highly nonlinear phase, once fast merging has started, the reconnection rate becomes effectively independent of the initially imposed driving timescale, 
and proceeds at a normalized rate close to $\sim 0.1$, in accordance with theoretical 
and simulation results on standard collisionless 
reconnection in preexisting current sheets \citep{Comisso2016, Goodbred2022}.

Our results provide quantitative estimates of the timescales of reconnection-driven flares in high-energy astrophysical environments, such as pulsar wind nebulae \citep{Abdo2011,Buehler2012},  black-hole accretion flows \citep{Abramowski2012, Rieger2018, GRAVITY2021, Algaba2024}, blazar jets \citep{Krawczynski2004}, or magnetospheres of magnetars \citep{Mereghetti2024}. 

Essentially, we find that reconnection of magnetically dominated flux tubes typically develops on a  timescale of a few light-crossing times. 
After the onset, the sheet fragments into  a chain of plasmoids. The first plasmoids are expelled from the sheet by the reconnection outflows. We expect a weak emission during the early stage of reconnection, which typically  lasts a few $R/c$. This is followed by a more violent flux-tube merger phase, featuring  a new generation of plasmoids, which corresponds  to the peak of the dissipation and  is largely insensitive to the details of the driving mechanism. The location and size of a high-energy astrophysical flare within the source is generally not well constrained by present observations. Flares with variability timescales as short as a few light-crossing times of the central object have been detected from accreting black holes \citep[e.g.,][]{Abramowski2012,Rieger2018,Algaba2024}. Such timescales are naturally explained by  the reconnection scenario investigated in the present paper. Future work will investigate particle acceleration and  the emerging electromagentic emission during all stages of the reconnection process, which can facilitate a more direct comparison with observations.

\begin{acknowledgments}
C.G.~thanks L. Sironi, D. Uzdensky and I.~Demidov for helpful discussions. D.G.~also thanks B.~\mbox{Ripperda} for useful discussions and
L.~\mbox{Sironi} for sharing his unpublished results on flux-tube mergers.
F.B.\ acknowledges support from the FED-tWIN programme (profile Prf-2020-004, project ``ENERGY''), issued by BELSPO, and from the FWO Junior Research Project G020224N granted by the Research Foundation -- Flanders (FWO). 
D.G.~is supported by the FWO Senior Postdoctoral  Fellowship 12B1424N.
L.C. acknowledges support from the NASA ATP award 80NSSC22K0667 and the National Science Foundation award PHY-2308944.
The resources and services used in this work were provided by the VSC (Flemish Supercomputer Center), 
funded by the FWO and the Flemish Government.
\end{acknowledgments}
{\software{\textsc{Tristan-MP v2} \citep{tristan}}}

\appendix

\section{Initial Magnetic-Field Profiles}
\label{app:profiles}

\begin{figure}[ht]
    \centering
    \includegraphics[width=0.5\textwidth]{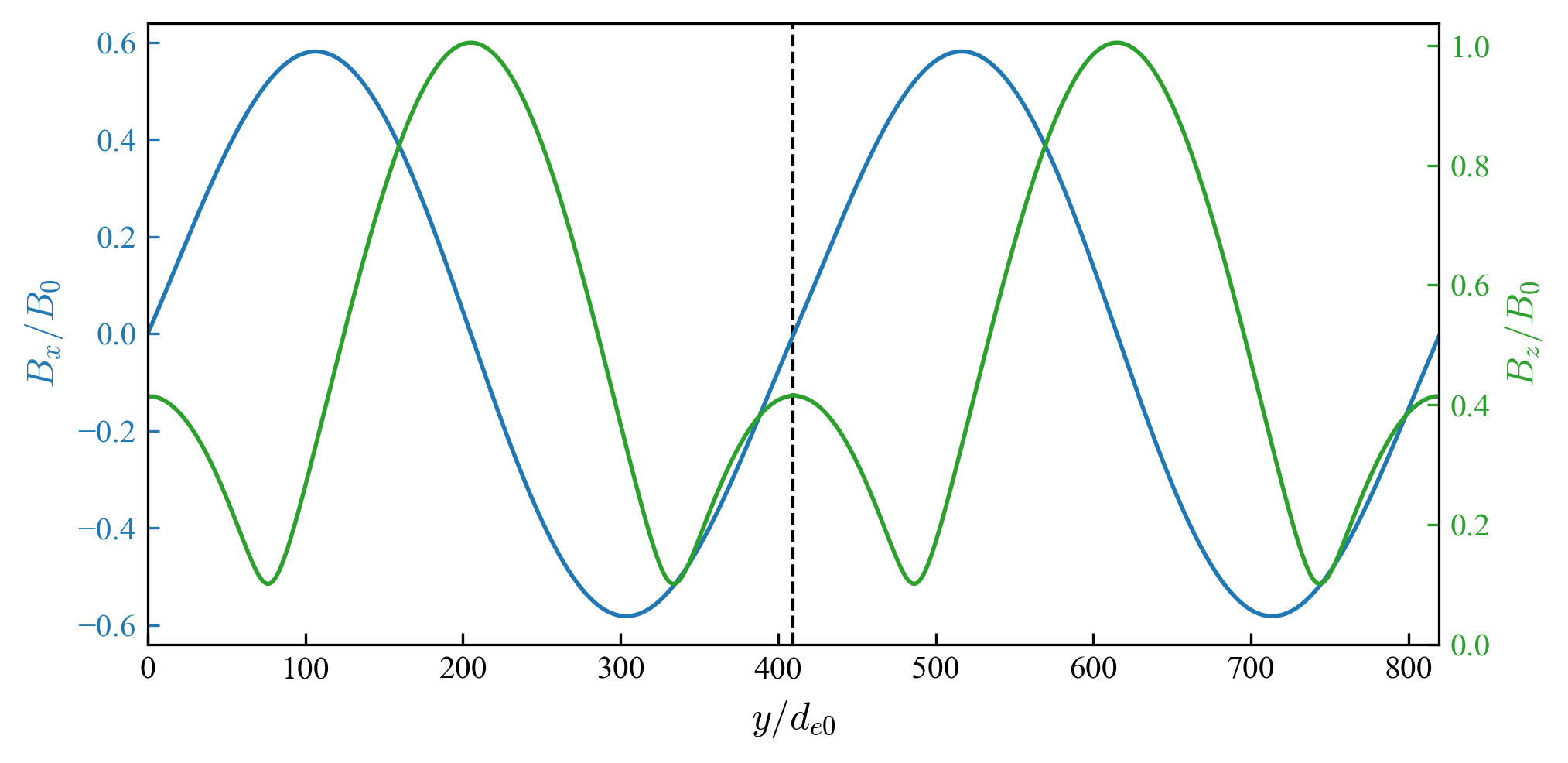}
    \caption{1D profiles of the initial magnetic-field components $B_x(x = 2R y)$ and $B_z(x = 2R y)$, for a flux-tube radius of $R/\dep=204$. The dashed vertical line indicates the position where the current sheet forms.}
    \label{fig:profiles}
\end{figure}

In Fig.~\ref{fig:profiles}, we show the initial 1D magnetic field profiles of $B_x$ and $B_z$ along the $y$ direction
at $x=2R$, with $R/\dep = 204$. The two flux tubes are centered at ${\bf r}_1 = (x_1, y_1) = (2R, R)$ and ${\bf r}_2 = (x_2, y_2) = (2R, 3R)$. The point of contact, where the current sheet eventually forms, is located at $y = 2 R$. The poloidal component $B_x$ vanishes at the center of the domain.  The guide field $B_z$ exhibits a nontrivial profile around the region where the sheet forms.

\section{Linear Tearing Modes}  
\label{app:lineargr}
\begin{figure}[ht]
\includegraphics[width=1\linewidth]{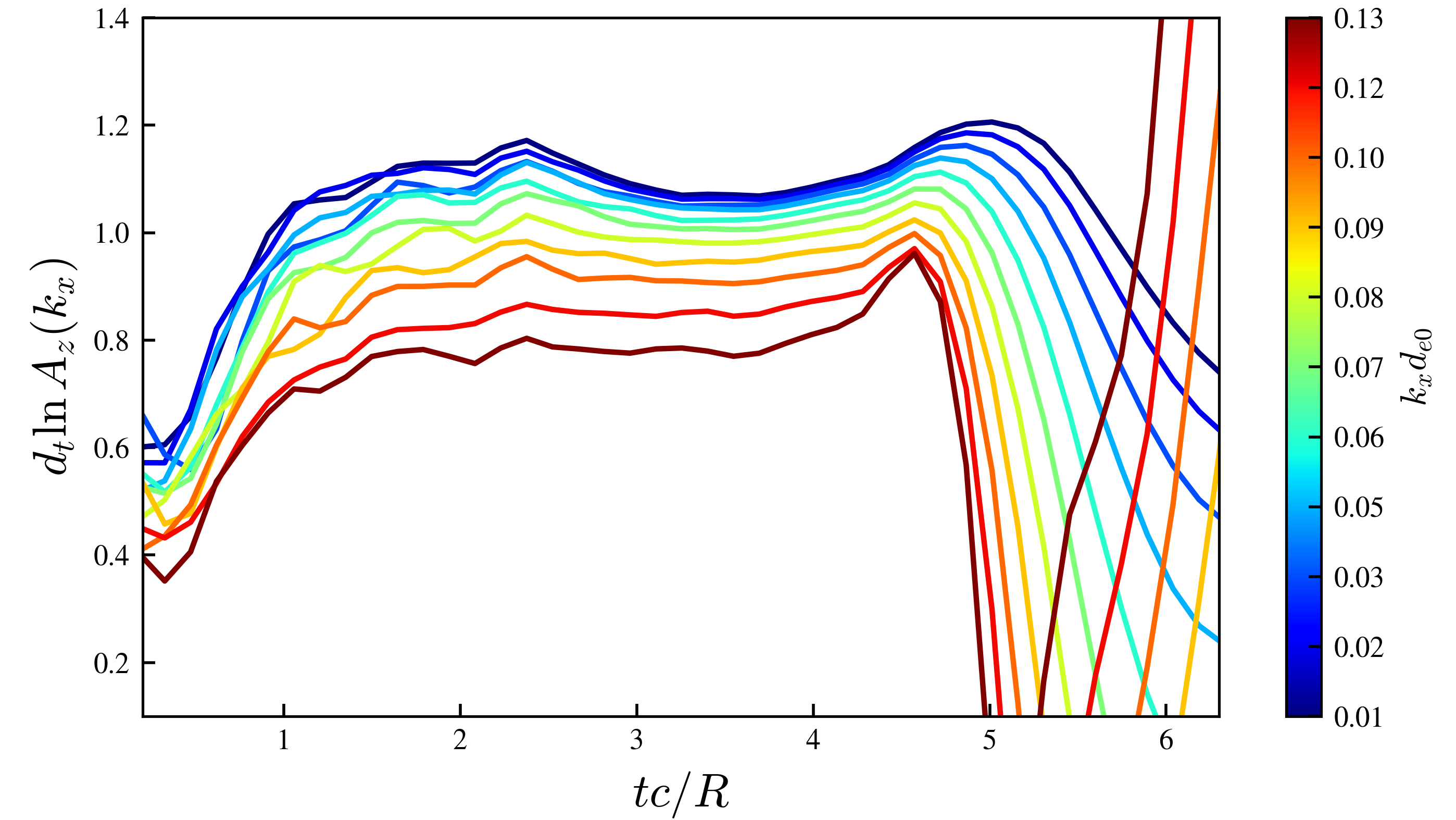}
    \caption{Growth rates of 10 tearing modes computed from the Fourier transform of $A_{z}(x)$, where $A_{z}$ is first averaged over $y$ inside the current sheet.
    The simulation parameters are $\sigma_{\rm in}=8$, $v_{\rm push}=0.02c$, $R/d_{e0}=204$.}
    \label{fig:lineargr}
\end{figure}

During Phase I, the current sheet thins while small-amplitude tearing-mode perturbations in the out-of-plane flux function $A_z$ grow exponentially. Only modes satisfying $k_x d_{e0} < 1$ are linearly unstable. Fig.~\ref{fig:lineargr} shows the time evolution of these linear tearing modes.
Here, $\sigma_{\rm in}=8$, $v_{\rm push}=0.02c$, and $R/d_{e0}=204$.

\begin{figure}[ht]
\includegraphics[width=1\linewidth]{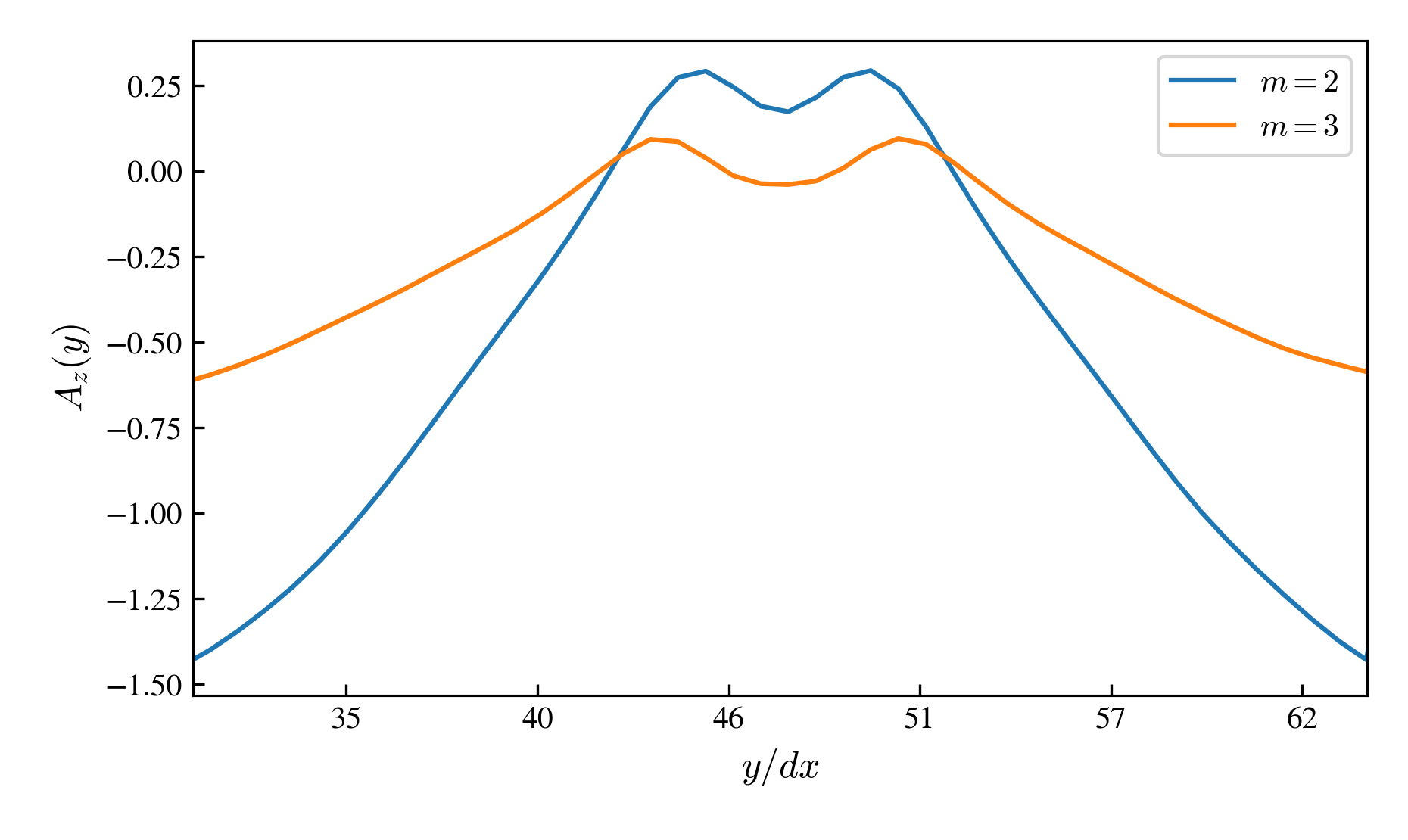}
    \caption{Flux function $A_z(y)$ for modes $m=2,3$, extracted from our simulation with $\sigma_0=50$, $\vp=0.02c$, and $R/\dep=408$.}
    \label{fig:appendix_eigen}
\end{figure}

We further verify the presence of linear tearing modes by extracting individual $k_x=m$ harmonics from the out‑of‑plane flux $A_z$ at $ t\approx 3.25 R/c$. After subtracting the $m=0$ mean, we reconstruct $A_z(y)$ for $m=2,3$. Figure~\ref{fig:appendix_eigen} displays each normalized eigenfunction, whose narrow peaks at the current‑sheet center span $\gtrsim10-15$ grid points. This shows that our mesh captures the tearing eigenmode reasonably well.

\section{Shock between the flux tubes}  
\label{app:shock}

The resulting shock‐heated inflow raises the upstream pressure and appears to slow down the reconnection. For the case with $\sigma_0=1$ and $v_{\rm push}=0.2\,c$, the fast‐merging Phase III lasts $\Delta t_{\rm merg}=6.7$ (normalized by $c/R$) instead of the nominal value of 2, illustrating the delayed collapse due to shock heating.

\begin{figure}[ht]
  \centering
  \includegraphics[width=\linewidth]{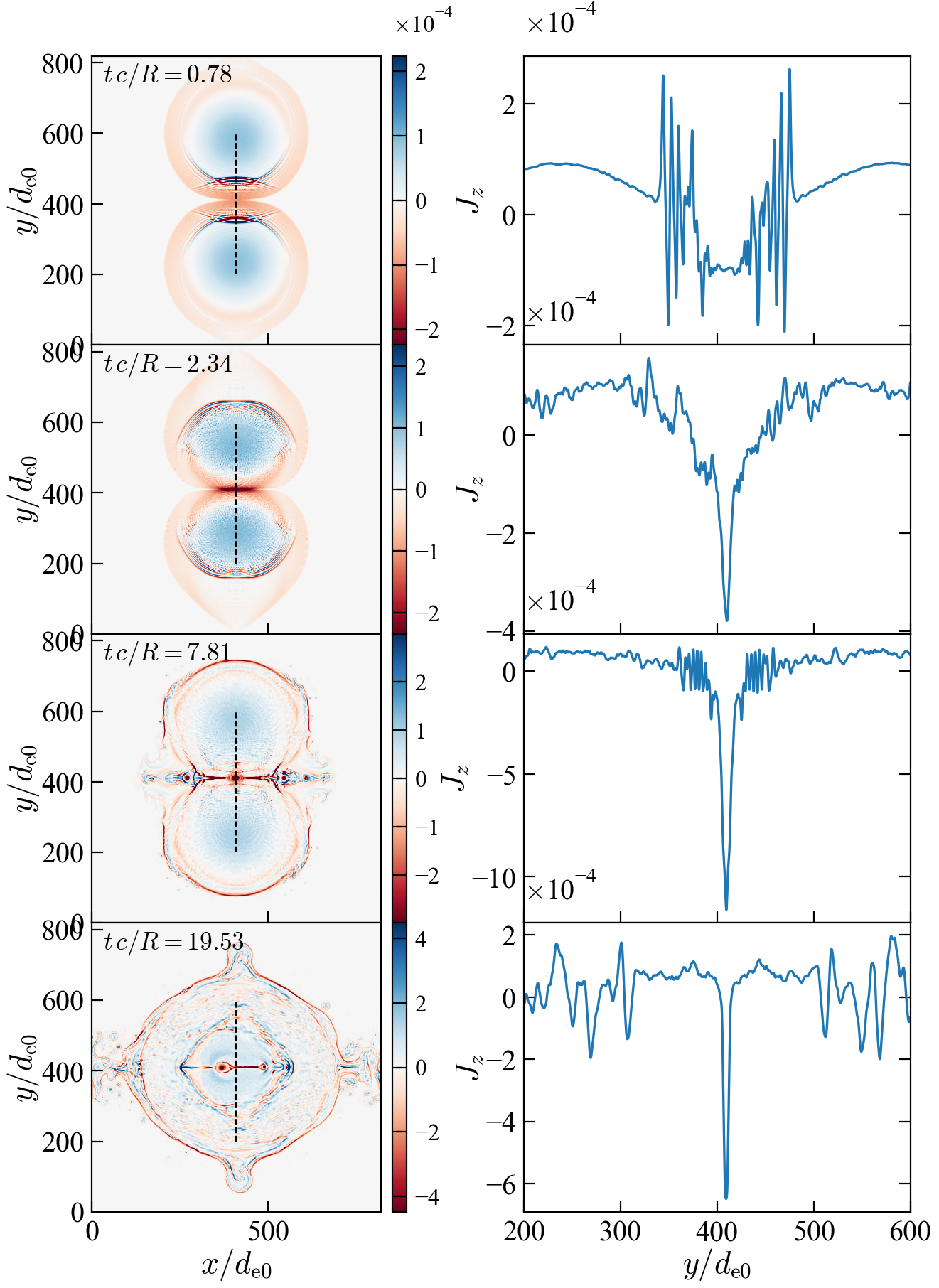}
  \caption{Current density and current-density cuts across the sheet at four different times for $\sigma_0=1$ and $v_{\rm push}=0.2\,c$ ($v_{\rm push}\gg c_{\rm ms}$).}
  \label{fig:jslices}
\end{figure}

\section{Dependence on spatial resolution and number of particles per cell}
\label{app:convergence}

\begin{figure}[ht]
    \includegraphics[width=1\linewidth]{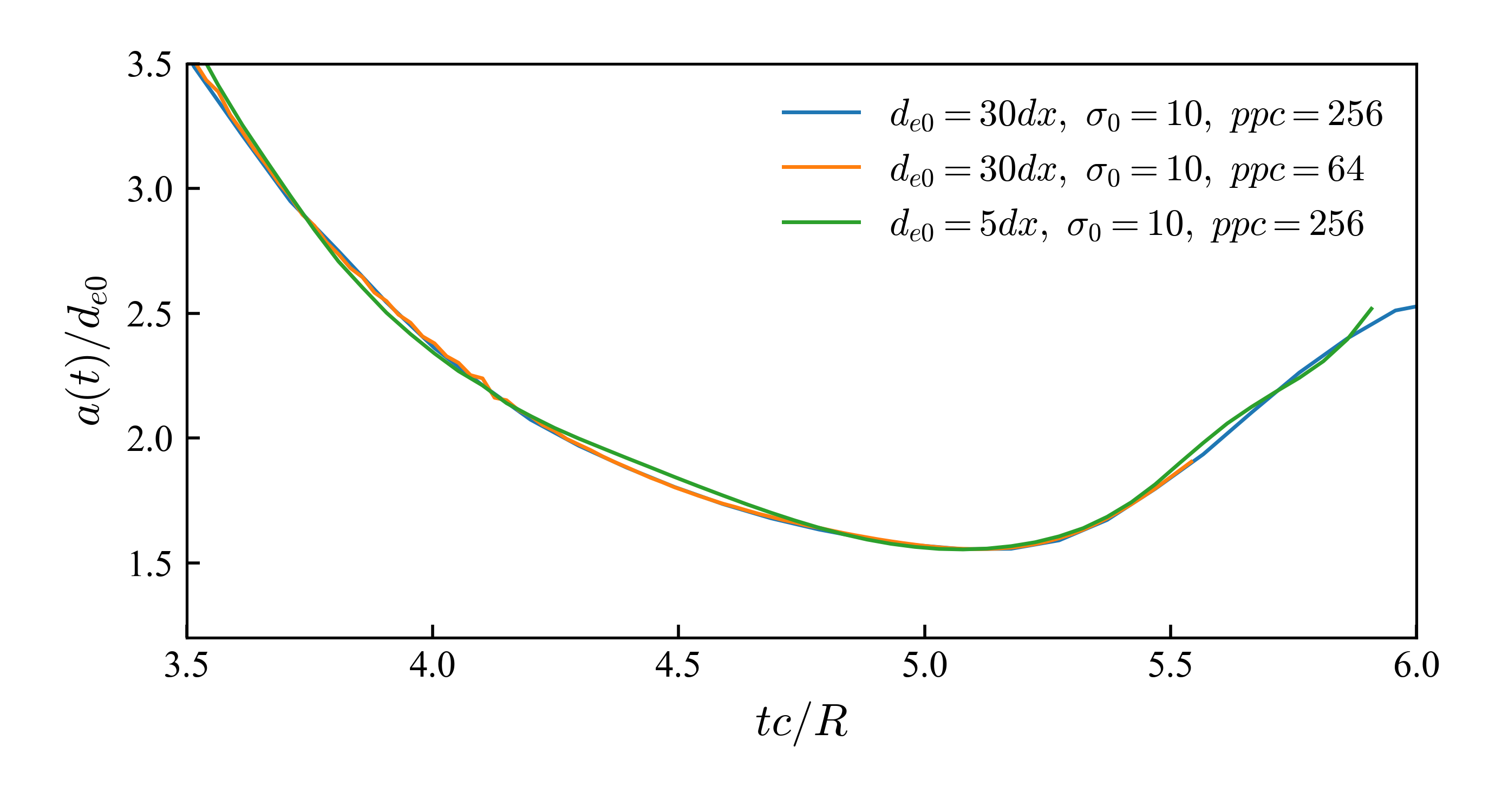} 
    \caption{Current‑sheet thickness for different spatial resolutions and  numbers of particle per cell at $v_{\rm push}=0.02c$.}
    \label{fig:convergence}
\end{figure}

In addition to the production runs, we conducted convergence tests at a fixed drive speed $v_{\rm push}=0.02c$ to verify the robustness of our results.  The tests vary both spatial resolution and particle count: a high‐resolution setup with $d_{e0}=30 dx$, $\rho_{e0}=9.48 d x$, and 256 ppc; the same grid spacing with 64 ppc; and a lower‐resolution case with $d_{e0}=5dx$, $\rho_{e0}=2.37dx$, and 256 ppc.  Figure \ref{fig:convergence} plots the time evolution of the sheet thickness in each run, demonstrating that our results are numerically robust.

\bibliographystyle{aasjournal} 
\bibliography{biblio}

\end{document}